\documentclass[letter,twocolumn,aps, showpacs]{revtex4}

\usepackage{graphicx}

\newcommand{\pderiv}[2]{\frac{\partial #1}{\partial #2}}
\newcommand{\pb}{p_{\rm b}}
\newcommand{\pd}{p_{\rm d}}
\newcommand{\pdc}{p_{\rm {d_c}}}

\newcommand{\df}[0]{d_{\rm f}}
\newcommand{\dfc}[0]{d_{{\rm f}_{\rm c}}}

\begin{document}

\title{Change in Order of Phase Transitions on Fractal Lattices}

\author{Alastair Windus}
\author{Henrik Jeldtoft Jensen}
 \email{h.jensen@imperial.ac.uk}
\affiliation{The Institute for Mathematical Sciences. 53 Prince's Gate, South
Kensington, London SW7 2PG}
\affiliation{Department of Mathematics, Imperial College London, South Kensington
Campus, London SW7 2AZ.}

\date{\today}

\begin{abstract}
We re-examine a population model which exhibits a continuous absorbing phase transition which belongs to directed percolation in 1+1 dimensions and a first order transition in 2+1 dimensions and above. Studying the model on fractal lattices, we examine at what fractal dimension $1<d_{\rm f}<2$, the change in order occurs. As well as commenting on the order of the transitions, we produce estimates for the critical points and, for continuous transitions, some critical exponents.
\end{abstract}


\maketitle

Continuous nonequilibrium phase transitions continue to be an area of great interest (see \cite{Hinrichsen_Non,Lubeck_Universal} for recent reviews). Perhaps the main reason for this is due to the observed universality that many such models display. As in equilibrium phase transitions,
models belonging to the same universality class have identical critical exponents
and the scaling functions become identical close to the critical point. By far the largest universality class of nonequilibrium phase transitions is directed percolation (DP). Indeed, it has been conjectured by Janssen and
Grassberger that all models with a scalar order-parameter that exhibit a continuous phase transition from an active state to a single absorbing state belong to the class \cite{Grassberger_On, Janssen_Non}.

\begin{figure}[tb]
        \centering\noindent
        {\small a)} \\
        \includegraphics[width=9cm]{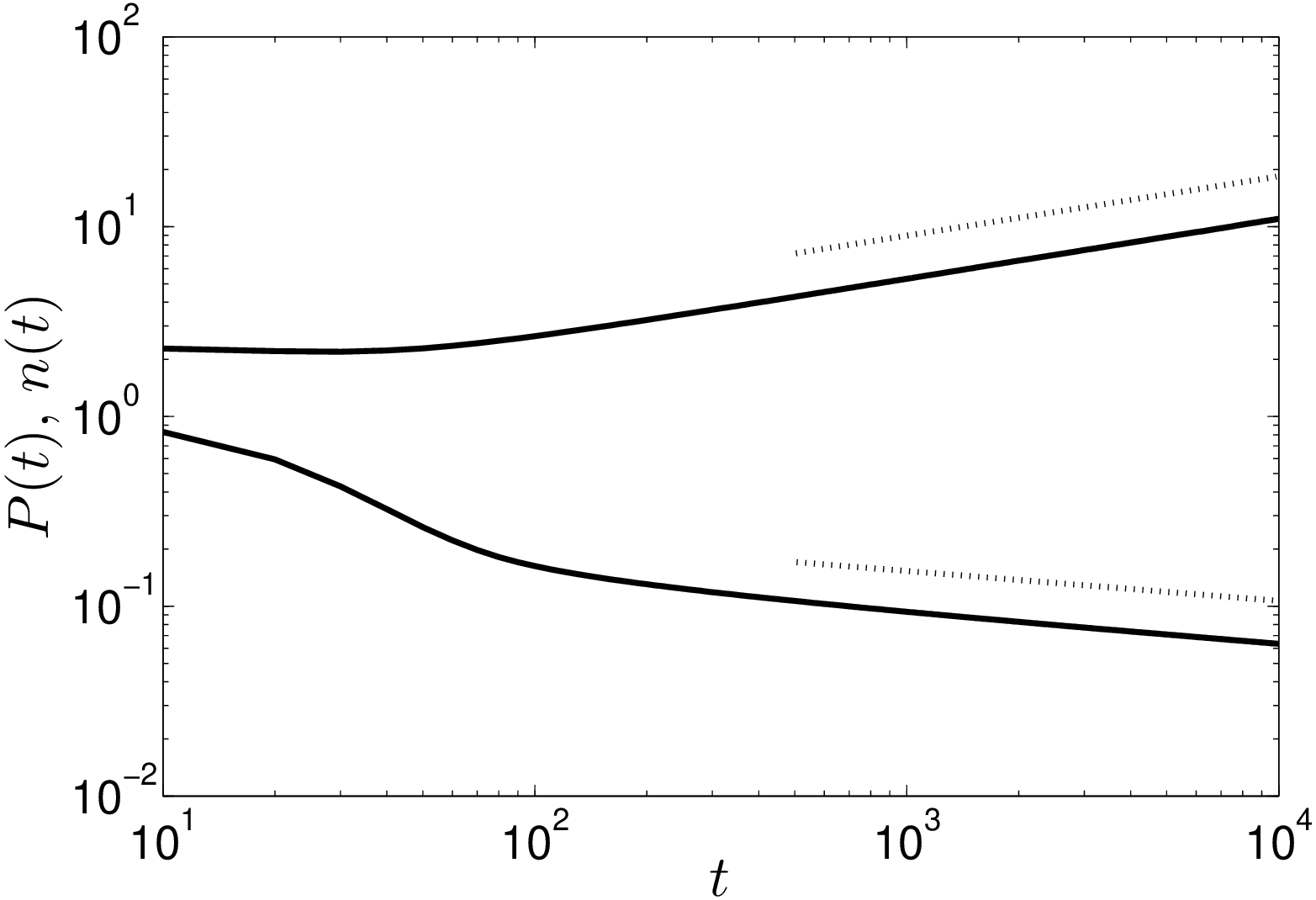} \\
        {\small b)} \\
        \includegraphics[width=9cm]{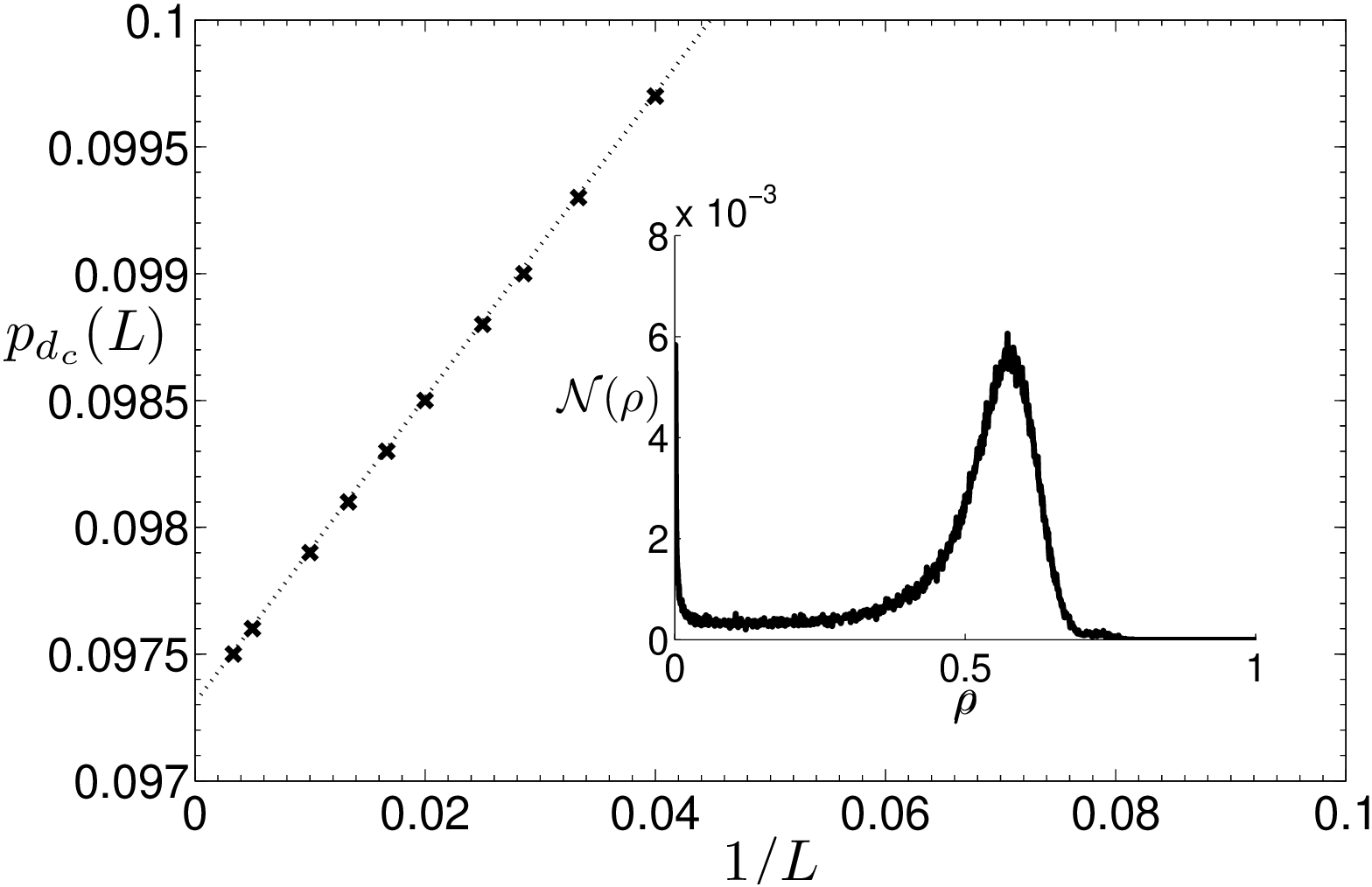}
        \caption{a) Power law behaviour at the continuous phase transition
        in 1+1 dimensions. The hashed lines show the DP values $\eta = 0.313686$
        and $\delta = 0.159464$ \cite{Jensen_Low}. b) The values of $\pdc(L)$
        with the hashed line of best fit which extrapolates the data as $L\rightarrow\infty$.
        The inset shows the double-peaked
        structure at the critical point for a finite $L$.}
        \label{F: Phase transitions}
        \end{figure}

Most of the research on phase transitions has been conducted on lattices of integer dimension. Motivated by the apparent dependence of critical exponents on
dimension, as well as topological features such as ramification and connectivity, work has also been carried out on lattices of fractal dimension $\df$
\cite{Jensen_Fractal,Madelbrot_FractalBook, Pruessner_Sierpinski}. In this paper, we examine a
model where the order of the phase transition changes from continuous to
first-order for $1<\df<2$. We investigate the model on fractal lattices to
examine at what dimension the change in order occurs $\dfc$. 
\\ \\
\textit{The model: }In a recent paper \cite{Windus}, we introduced a general population model with the reactions
        \begin{equation}
        2A+\phi\longrightarrow 3A, \quad A\longrightarrow\phi \quad\mbox{and}\quad
        A\phi\longleftrightarrow\phi A,
        \end{equation}
for an individual $A$. The model is simulated on $d$-dimensional square lattices of linear length $L$ where each site is either
occupied by a single particle (1) or is empty (0). A site is chosen at random. With probability $p_d$
the particle on an occupied site dies,  leaving the site empty. If the particle does not die, a nearest neighbour site is randomly chosen. If the neighbouring site is empty the particle moves there, otherwise, the particle reproduces with probability
$p_b$ producing a new particle on another randomly selected neighbouring
site, conditional on that site being empty. A time-step is defined
as the number of lattice sites $N=L^d$ and periodic boundary conditions are
used. Assuming spatial homogeneity, the mean field (MF) equation is easily found to be
        \begin{equation}
        \pderiv{\rho(t)}{t} = \pb(1-\pd)\rho(t)^2(1-\rho(t))-\pd\rho(t).         \end{equation}
This has three stationary states,
        \begin{equation} \label{steady states}
        \bar\rho_0 = 0, \quad \bar\rho_\pm=\frac{1}{2}\left(1\pm\sqrt{1-\frac{4\pd}{\pb(1-\pd)}}\right).
        \end{equation} 
For $\pd > \pb/(4+\pb)$, $\bar\rho_\pm$ are imaginary resulting in $\bar\rho_0$
being the only real stationary state. We therefore define the critical point
$\pd = \pdc = \pb/(4+\pb)$, marking the first-order phase transition between survival and extinction of the population. As we showed in our paper \cite{Windus}, from
monte carlo (MC) simulations, the model displays a continuous phase transition in 1+1 dimensions and belongs to DP. In higher integer
dimensions, consistent with the MF prediction, the model displays a first-order phase transition. The difference in order of the phase
transition between the MF and MC simulation results in 1+1 dimensions only
is likely due to the larger correlations present in lower dimensions. Here,
an individual is very likely to be able to find a mate on a neighbouring
site. In 1+1 dimensions then, the reproduction rate is more accurately proportional
to $\rho(1-\rho)$ rather than $\rho^2(1-\rho)$.
\\ \\
For the continuous phase transitions, we examined the population size $n(t)$ and survival probability $P(t)$ after beginning
the simulations from a single seed - two adjacent particles. We expect the
asymptotic power law behaviour \cite{Grassberger_Directed}
        \begin{equation}
        n(t) \;\bar\propto\; t^\eta \quad\mbox{and}\quad P(t) \;\bar\propto\;         t^{-\delta},
        \end{equation}
at the critical point. Such time-dependent simulations are known to be more efficient than steady state simulations which are also more prone to finite-size
effects \cite{Jensen_Universality,Grassberger_Directed}. Fig. \ref{F: Phase transitions} a) shows the results for the (1+1)-dimensional case                 
where the gradient in the log-log plots gives the DP exponents $\eta$ and $\delta$.
For the first-order phase transitions, we began our simulations from a fully-occupied
lattice and observed a double-peaked structure
in the histogram of population density $\mathcal{N}(\rho)$. Finding the value of $\pd$ which
equated the size of the two peaks at $\bar\rho_0$ and $\bar\rho_+$ gave the
value of $\pdc(L)$ due to the phase coexistence that occurs at first-order
transitions. Extrapolating the data as $L\rightarrow\infty$
gave a value for the critical point as shown in Fig. \ref{F: Phase transitions}
b).

Having described how we examined our model in integer dimensions, we turn
now to describing our methodology for the fractal lattices. 
\\ \\
\textit{Methodology and results: }Sierpinski carpets have been used widely to provide a generic model of fractals
to study physical phenomena in fractal dimensions (see for example \cite{Gefen,
 Gefen_1, Gefen_2, Gefen_3,Jensen_Fractal, Monceau}). They are formed by dividing a square
into $l^2$ sub-squares and removing ($l^2-N_{oc})$ of these sub-squares from the centre \cite{Mandelbrot_Fractals}. This procedure is then iterated on the remaining subsquares and
repeated $\kappa$ times. As $\kappa\rightarrow\infty$ a fractal structure,
denoted by $SC(l^2, N_{oc})$, is formed. For finite $\kappa$, however, we denote the structure $SC(l^2, N_{oc}, \kappa)$ which has $N=N^\kappa_{oc}$ sites. The Hausdorff fractal dimension of the structure as $\kappa\rightarrow\infty$ is then $\df=\ln(N_{oc})/\ln(l)$.
Using different values of $l$ and $N_{oc}$ enables us to use lattices of
different fractal dimension $\df$ where $1<\df<2$.

\begin{figure}[tb]
        \centering\noindent
        \includegraphics[width=8cm, trim=0cm 3cm 0cm 3cm, clip = true]{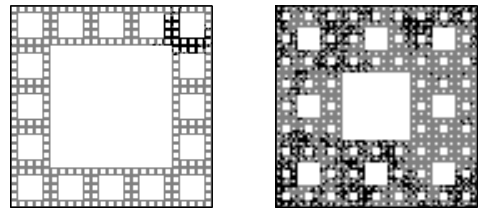}         \caption{Snapshots of the population on the fractal lattices $SC(5^2,16,3)$
        (left) and $SC(3^2,8,4)$. The left fractal with $\df = \log(16)/\log(5)\simeq
        17227$ was initiated with a single seed, whereas the right lattice
        with $\df = \log(8)/\log(3)\simeq 1.8928$ began from a fully-occupied
        lattice.}
        \label{F: FD Snapshots}
        \end{figure}

Using these fractal-dimensional lattice structures, we carried out our simulations
as before to determine the values for $\pdc$ and, where a continuous phase
transition occurred, the critical exponents.
For continuous phase transitions, we used time-dependent simulations, beginning
from a single-seed. Here, however, the position of
our initial seed may well affect the results. Two adjacent particles next
to a large hole would clearly be less likely to survive than two particles
surrounded by empty sites. We therefore randomly picked two adjacent sites
for each run and, as before, made an average over all runs. It is also clear that, unlike before, we are unable to begin the simulations
with a single-seed at the centre of the lattice due to the way in which the
lattices are constructed. Due to the random position
of this single-seed, we are not able to make the lattice sufficiently large
so that the particles never reach the boundary. As a compromise, we use very
large lattices and periodic boundary conditions. For the first-order
phase transitions, we examined the histograms of population density having
started from a fully-occupied lattice as described earlier. Example snapshots
of both of these approaches are shown in Fig. \ref{F: FD Snapshots}.

There are many problems associated with examining critical behaviour on fractal
lattices. Issues of boundary conditions and the initial location of the single
seed have already been mentioned. Further, there is the obvious difficulty
in the fact that finite lattices are not truly fractal. Indeed, for the histogram
approach, we are required to find $\pdc(N)$ for different lattice sizes which
we achieve by increasing $\kappa$. Such an approach, however, changes the structure of the lattice and can therefore affect the results. The effects
of these difficulties and inaccuracies are minimised here since we are primarily interested in the \textit{order} of the phase transition at the different dimensions. 

\begin{figure}[tb]
        \centering\noindent
        {\small a)} \\
        \includegraphics[width=9cm]{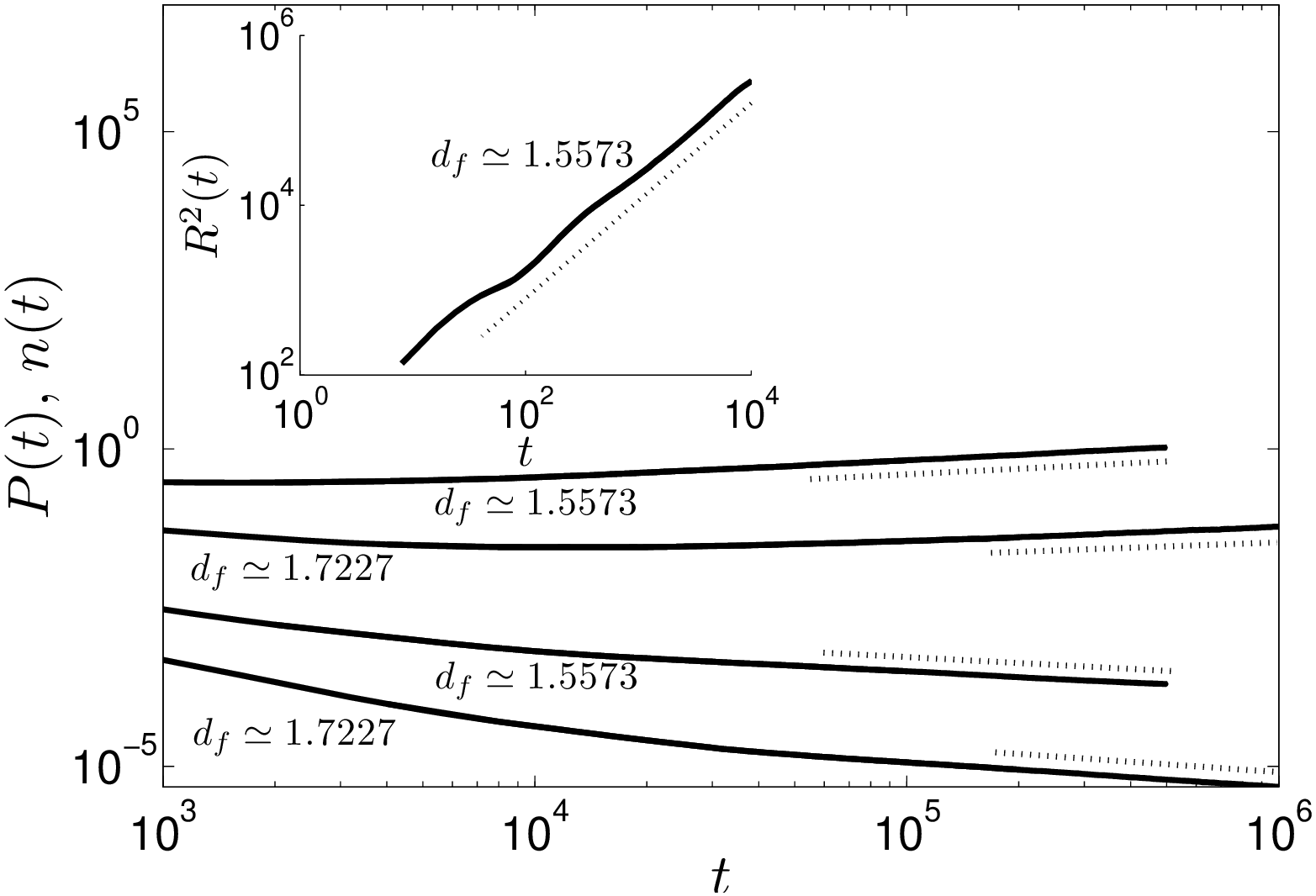}
        {\small b)} \\
        \includegraphics[width=9cm]{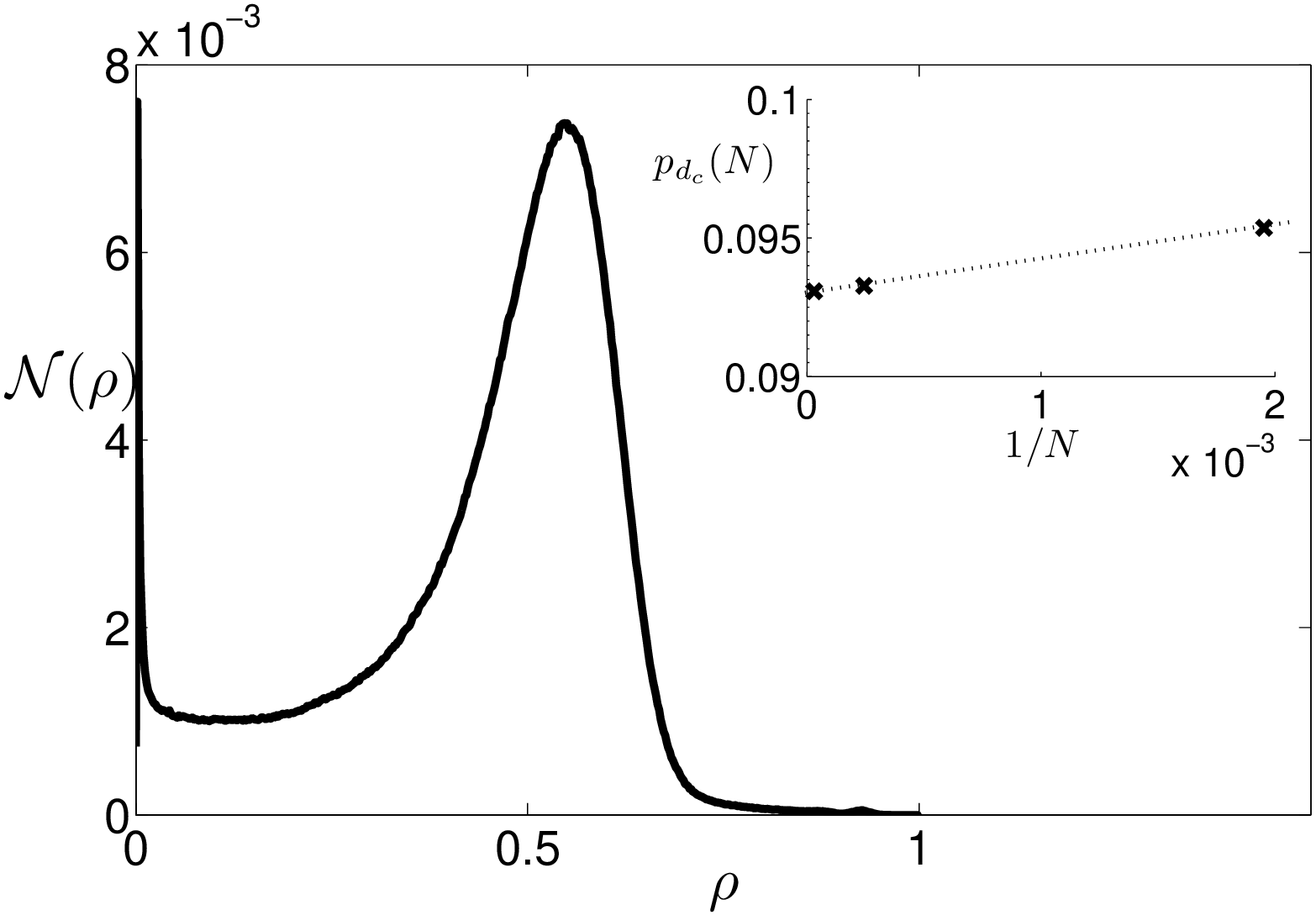}
        \caption{a) Plots of $n(t)$ and $P(t)$ at the critical point for         fractal dimensions 1.5573 and 1.7227. The hashed lines give the estimated         values for the exponents outlined in Tab. \ref{T: Fractal}. The inset
        shows $R^2(t)$ for $\df\simeq1.5573$ only. The hashed line has gradient
        $2/z$ where $z$ is obtained using the scaling relation (\ref{E: Scaling
        relation}) and the obtained values for $\eta$ and $\delta$. b) The
        double peaked histogram of population density indicating a first-order
        phase transition for $\df\simeq 1.8928$. The inset shows the predicted         values of $\pdc(N)$
        and an extrapolation of these results for $N\rightarrow\infty$.
        } 
        \label{F: PT}
        \end{figure}

For $\df \simeq 1.5573$ and $\df \simeq 1.7227$, we find continuous phase transitions shown by observing power law behaviour at the critical point.
Plots of both $n(t)$ and $P(t)$ are shown in Fig. \ref{F: PT} a).
For $\df \simeq 1.5573$, we used $SC(9^2,32, 4)$ which has just over $10^6$
sites. The data was obtained from over $1.6\times10^7$ independent runs. Comparing the
plots for this dimension with those for $d=1$, we observe significantly larger
corrections to scaling. Whereas power law behaviour was obtained after $\sim
10^2$ time steps for the $d=1$ case, for $\df\simeq1.5573$, we have to wait
$\sim10^4$ time steps and $\sim 10^5$ time steps for $\df\simeq1.7227$. For
this latter case, we therefore had to increase the number of time steps as well as the number of independent runs to $2.25\times10^8$. The simulations took over three months of computer time on Imperial College London's HPC \cite{HPC} for each parameter value. As the crossover in the order of transition
is approached, these corrections to scaling are likely to increase further
in size. This, then, represents a further challenge in obtaining accurate values for the critical points and exponents as $\df\rightarrow\dfc$.

From the other end of the interval $1<\df<2$, for $\df\simeq1.8928$, no power
law behaviour was observed, rather, the double-peaked structure of the histogram
of population density. To increase the size of our lattice, we increased
the value of $\kappa$ and found the value of $\pdc(N)$ for each case. Extrapolating
the results for infinite system size, we obtained an approximation for the
critical point as shown in Fig. \ref{F: PT} b).

For $\df\simeq1.7925$, 
        \begin{figure}[tb]
        \centering\noindent
        \includegraphics[width=9cm]{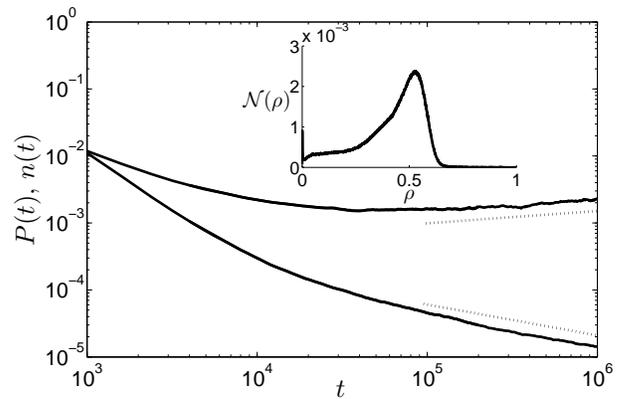}
        \caption{Possible power law behaviour for $\df\simeq1.7927$. The         inset shows the lack of the double-peaked structure.
        } 
        \label{F: PT2}
        \end{figure}
the histogram plots do not show the double-peaked
structure close to the critical point. Plots of $n(t)$ and $P(t)$ appear
to show power law behaviour, but with very large corrections to scaling.
The plots shown in Fig. \ref{F: PT2} required over $5\times 10^8$ runs,
taking over six months of computer time. On closer inspection, the values
of $\eta=0.19$ and $\delta=0.46$ that we obtained at this dimension appear
to be unlikely. Both vales are the wrong side of the DP values for $d=2$, indicating
that the plots actually show supercritical behaviour. Due to the large amounts
of time required to produce such plots, we predict that the phase transition
is continuous for $\df\simeq1.7925$ but are, now at least, unable to provide
strong evidence for this.

A summary of our results at the different dimensions are outlined in Tab.
\ref{T: Fractal}. 
        \begin{table*}[tb]
        \centering\noindent
        \begin{tabular}{cccccc}
        \hline\hline
        $\df$ & $SC(l^2, N_{oc}, \kappa)$ & Order of   & $\pdc$ & $\delta$
        & $\eta$\\
        & & phase transition &  & (if applicable) & (if applicable)
        \\
        \hline
        1& & continuous  & 0.071754 & 0.160 & 0.313
        \\
        $\log(32)/\log(9)\simeq 1.5573$ & $SC(9^2,32, 4)$ & continuous &         0.08553         & 0.30 & 0.29  \\
        $\log(16)/\log(5)\simeq 1.7227$ & $SC(5^2,16, 6)$ &continuous & 0.08819         & 0.39 & 0.24 \\
        $\log(12)/\log(4)\simeq 1.7925$ & $SC(4^2,12,6)$ & ? & ? & ? & ? \\
        $\log(8)/\log(3)\simeq 1.8928$ & $SC(3^2,8,3/4/5)$ &first-order &         0.093         & n/a & n/a \\
        2& & first-order & 0.0973 & n/a & n/a \\
        \hline\hline
        \end{tabular}
        \caption{Summary of the results. The data for $d=1,2$ was obtained
        in \cite{Windus}. Due to the large corrections to scaling, we were
        unable to provide strong evidence for the order of the phase transition
        for $\df\simeq 1.7925$. While we predict it to be continuous, we
        leave this row with question marks due to our uncertainty.}
        \label{T: Fractal}
        \end{table*}
For the continuous phase transitions, a generally accepted
approach of obtaining more accurate values for the critical exponents is
by examining the local slopes (see for example \cite{Grassberger_Directed}). For such an approach we plot, for example, $\delta(t)$ against $1/t$ where
        \begin{equation}  \label{local slope}
        -\delta(t) = \frac{\ln\left[P(t)/P(t/m)\right]}{\ln(m)},
        \end{equation}
and $m$  is the local range over which the slope is measured. Due to the
previously mentioned large corrections to scaling we observed, such an approach
did not yield accurate results. For the method to work properly,
the number of time steps over which the data was collected would have to
be raised significantly, increasing the required computer time further. The values for the exponents that we obtained then were estimated
only by measurement of the gradient of the above plots and are therefore
accurate to only $\pm 0.005$.

A check for the consistency of the exponents can be obtained by the scaling
relations. From the scaling relation \cite{Grassberger_Directed} 
        \begin{equation} \label{E: Scaling relation}
        \eta+2\delta=d/z,
        \end{equation}
we can check
our values of $\eta$ and $\delta$ using the fractal dimension for $d$ and
obtaining the dynamical exponent $z=\nu_\parallel/\nu_\perp$. This exponent can be obtained by examining the
mean square distance of spread from the initial seed of the population $R^2(t)$
averaged over surviving runs only. At the critical point, we expect the asymptotic
power law behavior $R^2(t)\;\bar\propto\;t^{2/z}$ \cite{Grassberger_Directed}. For $\df=1.5573$ we show the results in the inset of Fig. \ref{F: PT}
a) where we plot $R^2(t)$ along with $t^{2/z}$ using the value of $z$ obtained from the scaling
relation (\ref{E: Scaling relation}) given by the obtained values for $\eta$ and $\delta$. We see very
good agreement between the exponents. For larger dimensions, however, the
crossover effects render, in particular $R^2(t)$, inaccurate. Oscillations
in $R^2(t)$, perhaps due to the fractal structure, delay the power-law behaviour
to large values of $t$ that are impractical to simulate.
\\ \\
\textit{Remarks: }Having examined our model in different fractal dimensions, we found that
the dimension at which the order of the phase transition changes is in the
range $1.7227 < \dfc < 1.8928$. We predict the lower bound of this range to be more accurately given by $1.7927$ but offer no firm proof. We predict
that below $\dfc$ the larger correlations between the particles ensure that
a continuous phase transition is observed.

This present study was limited by the available computer power (over two years of processing time was required in total). Due to the large corrections to scaling, it was difficult to obtain truly accurate results for the critical points and exponents in the continuous
regime. Given more computer time, we would be able to run the simulations over a greater number of time steps to increase the accuracy. Further, it would also be interesting to examine if the values of the critical points and exponents depend on the type of fractal lattices used. 


\begin{thebibliography}{18}
\expandafter\ifx\csname natexlab\endcsname\relax\def\natexlab#1{#1}\fi
\expandafter\ifx\csname bibnamefont\endcsname\relax
  \def\bibnamefont#1{#1}\fi
\expandafter\ifx\csname bibfnamefont\endcsname\relax
  \def\bibfnamefont#1{#1}\fi
\expandafter\ifx\csname citenamefont\endcsname\relax
  \def\citenamefont#1{#1}\fi
\expandafter\ifx\csname url\endcsname\relax
  \def\url#1{\texttt{#1}}\fi
\expandafter\ifx\csname urlprefix\endcsname\relax\def\urlprefix{URL }\fi
\providecommand{\bibinfo}[2]{#2}
\providecommand{\eprint}[2][]{\url{#2}}

\bibitem[{\citenamefont{Hinrichsen}(2000)}]{Hinrichsen_Non}
\bibinfo{author}{\bibfnamefont{H.}~\bibnamefont{Hinrichsen}},
  \bibinfo{journal}{Adv. Phys.} \textbf{\bibinfo{volume}{49}},
  \bibinfo{pages}{815} (\bibinfo{year}{2000}).

\bibitem[{\citenamefont{L\"ubeck}(2004)}]{Lubeck_Universal}
\bibinfo{author}{\bibfnamefont{S.}~\bibnamefont{L\"ubeck}},
  \bibinfo{journal}{Int. J. Mod. Phys. B} \textbf{\bibinfo{volume}{18}},
  \bibinfo{pages}{3977} (\bibinfo{year}{2004}).

\bibitem[{\citenamefont{Grassberger}(1982)}]{Grassberger_On}
\bibinfo{author}{\bibfnamefont{P.}~\bibnamefont{Grassberger}},
  \bibinfo{journal}{Z. Phys. B} \textbf{\bibinfo{volume}{47}},
  \bibinfo{pages}{365} (\bibinfo{year}{1982}).

\bibitem[{\citenamefont{Janssen}(1981)}]{Janssen_Non}
\bibinfo{author}{\bibfnamefont{H.~K.} \bibnamefont{Janssen}},
  \bibinfo{journal}{Z. Phys. B} \textbf{\bibinfo{volume}{42}},
  \bibinfo{pages}{151} (\bibinfo{year}{1981}).

\bibitem[{\citenamefont{Jensen}(1991{\natexlab{a}})}]{Jensen_Fractal}
\bibinfo{author}{\bibfnamefont{I.}~\bibnamefont{Jensen}}, \bibinfo{journal}{J.
  Phys. A: Math. Gen.} \textbf{\bibinfo{volume}{24}}, \bibinfo{pages}{L1111}
  (\bibinfo{year}{1991}{\natexlab{a}}).

\bibitem[{\citenamefont{Mandelbrot}(1983)}]{Madelbrot_FractalBook}
\bibinfo{author}{\bibfnamefont{B.}~\bibnamefont{Mandelbrot}},
  \emph{\bibinfo{title}{The Fractal Geometry of Nature}}
  (\bibinfo{publisher}{W.H. Freeman and Company, New York},
  \bibinfo{year}{1983}).

\bibitem[{\citenamefont{Pruessner et~al.}(2001)\citenamefont{Pruessner, Loison,
  and Schotte}}]{Pruessner_Sierpinski}
\bibinfo{author}{\bibfnamefont{G.}~\bibnamefont{Pruessner}},
  \bibinfo{author}{\bibfnamefont{D.}~\bibnamefont{Loison}}, \bibnamefont{and}
  \bibinfo{author}{\bibfnamefont{K.D.}~\bibnamefont{Schotte}},
  \bibinfo{journal}{Phys. Rev. B} \textbf{\bibinfo{volume}{64}},
  \bibinfo{pages}{134414} (\bibinfo{year}{2001}).

\bibitem[{\citenamefont{Windus and Jensen}(2007)}]{Windus}
\bibinfo{author}{\bibfnamefont{A.}~\bibnamefont{Windus}} \bibnamefont{and}
  \bibinfo{author}{\bibfnamefont{H.}~\bibnamefont{Jensen}},
  \bibinfo{journal}{J. Phys. A: Math. Theor.} \textbf{\bibinfo{volume}{40}},
  \bibinfo{pages}{2287} (\bibinfo{year}{2007}).

\bibitem[{\citenamefont{Grassberger}(1989)}]{Grassberger_Directed}
\bibinfo{author}{\bibfnamefont{P.}~\bibnamefont{Grassberger}},
  \bibinfo{journal}{J. Phys. A: Math. Gen.} \textbf{\bibinfo{volume}{22}},
  \bibinfo{pages}{3673} (\bibinfo{year}{1989}).

\bibitem[{\citenamefont{Jensen}(1991{\natexlab{b}})}]{Jensen_Universality}
\bibinfo{author}{\bibfnamefont{I.}~\bibnamefont{Jensen}},
  \bibinfo{journal}{Phys. Rev. A} \textbf{\bibinfo{volume}{43}},
  \bibinfo{pages}{3187} (\bibinfo{year}{1991}{\natexlab{b}}).

\bibitem[{\citenamefont{Jensen}(1999)}]{Jensen_Low}
\bibinfo{author}{\bibfnamefont{I.}~\bibnamefont{Jensen}}, \bibinfo{journal}{J.
  Phys. A: Math. Gen.} \textbf{\bibinfo{volume}{32}}, \bibinfo{pages}{5233 }
  (\bibinfo{year}{1999}).

\bibitem[{\citenamefont{Gefen et~al.}(1980)\citenamefont{Gefen, Mandelbrot, and
  Aharony}}]{Gefen}
\bibinfo{author}{\bibfnamefont{Y.}~\bibnamefont{Gefen}},
  \bibinfo{author}{\bibfnamefont{B.B.}~\bibnamefont{Mandelbrot}},
  \bibnamefont{and} \bibinfo{author}{\bibfnamefont{A.}~\bibnamefont{Aharony}},
  \bibinfo{journal}{Phys. Rev. Lett.} \textbf{\bibinfo{volume}{45}},
  \bibinfo{pages}{855} (\bibinfo{year}{1980}).

\bibitem[{\citenamefont{Gefen et~al.}(1983)\citenamefont{Gefen, Aharony, and
  Mandelbrot}}]{Gefen_1}
\bibinfo{author}{\bibfnamefont{Y.}~\bibnamefont{Gefen}},
  \bibinfo{author}{\bibfnamefont{A.}~\bibnamefont{Aharony}}, \bibnamefont{and}
  \bibinfo{author}{\bibfnamefont{B.}~\bibnamefont{Mandelbrot}},
  \bibinfo{journal}{J. Phys. A: Math. Gen.} \textbf{\bibinfo{volume}{16}},
  \bibinfo{pages}{1267} (\bibinfo{year}{1983}).

\bibitem[{\citenamefont{Gefen et~al.}(1984{\natexlab{a}})\citenamefont{Gefen,
  Aharony, Shapir, and Mandelbrot}}]{Gefen_2}
\bibinfo{author}{\bibfnamefont{Y.}~\bibnamefont{Gefen}},
  \bibinfo{author}{\bibfnamefont{A.}~\bibnamefont{Aharony}},
  \bibinfo{author}{\bibfnamefont{Y.}~\bibnamefont{Shapir}}, \bibnamefont{and}
  \bibinfo{author}{\bibfnamefont{B.}~\bibnamefont{Mandelbrot}},
  \bibinfo{journal}{J. Phys. A: Math. Gen.} \textbf{\bibinfo{volume}{17}},
  \bibinfo{pages}{435} (\bibinfo{year}{1984}{\natexlab{a}}).

\bibitem[{\citenamefont{Gefen et~al.}(1984{\natexlab{b}})\citenamefont{Gefen,
  Aharony, and B.B.}}]{Gefen_3}
\bibinfo{author}{\bibfnamefont{Y.}~\bibnamefont{Gefen}},
  \bibinfo{author}{\bibfnamefont{A.}~\bibnamefont{Aharony}}, \bibnamefont{and}
  \bibinfo{author}{\bibfnamefont{B.}~\bibnamefont{Mandelbrot}}, \bibinfo{journal}{J.
  Phys. A: Math Gen.} \textbf{\bibinfo{volume}{17}}, \bibinfo{pages}{1277}
  (\bibinfo{year}{1984}{\natexlab{b}}).

\bibitem[{\citenamefont{Monceau and Hsiao}(2004)}]{Monceau}
\bibinfo{author}{\bibfnamefont{P.}~\bibnamefont{Monceau}} \bibnamefont{and}
  \bibinfo{author}{\bibfnamefont{P.Y.} \bibnamefont{Hsiao}},
  \bibinfo{journal}{Phys. Lett. A} \textbf{\bibinfo{volume}{332}},
  \bibinfo{pages}{310} (\bibinfo{year}{2004}).

\bibitem[{\citenamefont{Mandelbrot}(1977)}]{Mandelbrot_Fractals}
\bibinfo{author}{\bibfnamefont{B.}~\bibnamefont{Mandelbrot}},
  \emph{\bibinfo{title}{Fractals: Form, Chance and Dimension}}
  (\bibinfo{publisher}{Freeman, New York}, \bibinfo{year}{1977}).

\bibitem[{HPC()}]{HPC}
\emph{\bibinfo{title}{\url{www.imperial.ac.uk/ict/services/teachingandr%
esearchservices/highperformancecomputing}}}.

\end{thebibliography}

\end{document}